\documentclass[aps,showpacs,preprint,superscriptaddress]{revtex4}
\usepackage{graphicx}
\usepackage{subfigure}
\usepackage{float}
\usepackage{amsmath}
\usepackage{amsfonts}
\usepackage{color}
\usepackage{bm}
\usepackage{bbm}
\usepackage{txfonts}
\usepackage{array}
\usepackage{amssymb}

\begin{document}
\title{Thomson backscattering in combined two laser and magnetic field}
\author{Zhijing Chen}
\affiliation{Key Laboratory of Beam Technology of the Ministry of Education, and College of Nuclear Science and Technology, Beijing Normal University, Beijing 100875, China}
\author{Li Zhao}
\affiliation{Key Laboratory of Beam Technology of the Ministry of Education, and College of Nuclear Science and Technology, Beijing Normal University, Beijing 100875, China}
\author{Chun Jiang}
\affiliation{Key Laboratory of Beam Technology of the Ministry of Education, and College of Nuclear Science and Technology, Beijing Normal University, Beijing 100875, China}
\author{Haibo Sang}
\affiliation{Key Laboratory of Beam Technology of the Ministry of Education, and College of Nuclear Science and Technology, Beijing Normal University, Beijing 100875, China}
\author{Bai-Song Xie \footnote{corresponding author: bsxie@bnu.edu.cn}}
\affiliation{Key Laboratory of Beam Technology of the Ministry of Education,
and College of Nuclear Science and Technology, Beijing Normal University, Beijing 100875, China}
\affiliation{Beijing Radiation Center, Beijing 100875, China}

\date{\today}
\begin{abstract}
The Thomson backscattering of an electron moving in combined fields is studied by a dynamically assisted mechanism. The combined fields are composed of two co-propagating laser fields and a magnetic field, where the first laser field is strong and low-frequency while the second is weak and high-frequency, relatively. The dependence of fundamental frequency of emission on the ratio of incident laser high-to-low frequency is presented and the spectrum of backscattering is obtained. It is found that, with a magnetic field, the peak of the spectrum and the corresponding radiation frequency are significantly larger in case of two-laser than that in case of only one laser. They are also improved obviously as the frequency of the weak laser field. Another finding is the nonlinear correlation between the emission intensity of the backscattering and the intensity of the weak laser field. These results provide a new possibility to adjust and control the spectrum by changing the ratios of frequency and intensity of the two laser fields.
\end{abstract}
\maketitle

\section{Introduction}

Thomson backscattering is widely used to create x- and $\gamma$-ray. During the past twenty years, several studies about Thomson backscattering have revealed how the spectrum depend on the laser field\cite{Salamin1996Harmonic,Salamin1997Ponderomotive,Salamin1997Harmonic} and external magnetic field\cite{Faisal1999Electron,Salamin2008Relativistic}.

At the beginning of this century, He \textit{et al} present that the spectra of Thomson scattering does not occur at integer multiples of the laser frequency and get a simple scaling law for the spectra to the electron energy in a linearly polarized plane wave\cite{He2002Phase,He2003Backscattering}.
Actually the perspective of the laser-magnetic resonance acceleration \cite{Gupta2005Electron,Singh2006Acceleration,Singh2005Electron,He2003Ponderomotive,Liu2004Resonance} has shown that the relativistic cyclotron motion will occur in transverse direction under an external magnetic field. There are a lot of advantages of electron resonance acceleration in laser field assisted by the external magnetic field\cite{Li2002Nonlinear}.

It is demonstrated that, in the plane of the electron cyclotron orbit, the electron will emit radiation at high harmonics of the cyclotron frequency\cite{Yu2002Electron}. Furthermore, the cyclotron-resonance condition is met when the cyclotron frequency approaches to the laser frequency, and the helically periodic motion of electrons would result in that the fundamental frequency of harmonic emission is neither the laser frequency nor the cyclotron frequency for the backscattered Thomson spectra\cite{He2002Phase}.

Since the cyclotron-resonance motion in the combined laser and magnetic fields will affect the frequency of emission.
In our previous work, Thomson scattering in combined fields with a circularly\cite{Fu2016Scale} or elliptically\cite{Jiang2017Thomson} polarized laser field and a strong magnetic field has been studied in detail. The scale invariance and scaling law for the laser intensity and initial axial momentum are found\cite{Fu2016Scale}. Another interesting finding is a new way to produce THz emission by using Thomson backscattering\cite{Jiang2017Thomson}.

In the past two years, strong nonlinear characteristics of Thomson scattering have been reported. With nonlinearity, the shape of backscattering spectrum can be controlled by proper laser chirping. The results allow prediction of spectral form and narrowing bandwidth together with high efficiency\cite{Ghbregziabher2013Spectral,S2016Controlling}.

The case of Thomson scattering using two counter-propagating laser pulses has also been reported\cite{Liu2010Generation}. It is found that the radiation power of electron's Thomson scattering can be significantly enhanced in comparison with that obtained by only using one of these two pulses, both in the case of linear and circular polarization. Moreover, the Compton scattering using two different wavelength lasers has been considered\cite{Sakai2011Harmonic}. It has been proposed that it can provide a new way for controlling scattered photon energy distributions. The electron trajectory and photon energy have been calculated under different ratios of the two lasers' intensity. And the best ratio of intensity can be found through comparison.

It is worthy saying a few words about that when the second laser field is introduced, the radiation spectra exhibit very complicated phenomena and the radiation is enhanced greatly. The idea is very similar to the dynamically assisted mechanism in the study of electron-position pair creation \cite{Sch¨¹tzhold2008Dynamically,Li2014Dynamically,Aleksandrov2018Dynamically,Akal2019Simulating} where a weak but rapidly varying field is added to the original strong but slowly varying electric field. It leads to the strong nonlinear interference in particles momenta spectra and then the enhanced pair production will be achieved. Interestingly this physical process and mechanism seems to exist in many different research area such as high-harmonic generation, above-threshold ionization and so on. Motivated by this mechanism, in this letter, we will reveal that
the electronic nonlinear dynamics is enlarged assisted by the second field addition in present research configurations of two laser fields plus magnetic field. And this strong nonlinear dynamically assisted mechanism can not only cause the high oscillations but also enhance the intensity as well broaden the range in the Thomson backscattering spectra.

For simplicity as well the convenient analytical treatment the laser field is assumed as plane-wave form. The analytic results of momentum of the electron are got in this case.
Our work will focus on three aspects. First, the behavior of electron's motion is examined and calculated. And the influence of magnetic field is discussed. Second, the fundamental frequency of harmonic radiation is calculated and its parameters dependence are obtained and discussed, which shows some nonlinear phenomena. Third, the radiation spectrum in the combined fields is presented.
Our research shows that the dependence of the peak intensity and the corresponding frequency of the radiation spectrum on the ratio of two lasers exhibits some complex phenomena.
We believe that these results not only provide a much deeper understanding of Thomson backscattering, but also give a new method to control or/and adjust the spectrum by using a second laser field.

By the way there is also a deficiency of current method in this study, as in cases of   Refs.\cite{Fu2016Scale,Jiang2017Thomson,Sakai2011Harmonic}, that the recoil of the electron which is caused by the emission of a single photon and the radiation backreaction which occurs due to the energy loss during emission are neglected. This ignorance is valid and reasonable, which is checked below in the section of numerical results.

\section{Theoretical framework}\label{basic}

The formalism of Thomson scattering spectra can be derived in terms of equations of an electron (with mass $m$ and charge $-e$) moving in combined laser and magnetic field.
The magnetic field $B_{0}$ is parallel to the laser propagating direction ($z$ direction). The phase of the laser field is denoted as $\eta={\omega}_{0}t-\textit{\textbf{k}}\cdot\textit{\textbf{r}}$, where ${\omega}_{0}$ is the laser frequency, $t$ is the time, $\textit{\textbf{k}}$ and $\textit{\textbf{r}}$ are the wave vector and electron displacement vector, respectively.

We assume that the phase has an initial value $\eta_{in}=-z_{in}$ at $t=0$.
Similar to the techniques in \cite{Fu2016Scale}, we obtain the $m$-th spectrum of the harmonic radiation in unit of erg/s per unit solid angle for backscattering as follows
\begin{equation}\label{eqbasic1}
\frac{d^2 I_m}{d\Omega dt}=\frac{e^2}{4\pi^2c}\frac{1}{\varsigma^2}(m\omega_{1})^2(|\textit{\textbf{F}}_{mx}|^2+|\textit{\textbf{F}}_{my}|^2),\\
\end{equation}
where
\begin{equation}\label{eqbasic2}
\textit{\textbf{F}}_{mx,my}=\omega_{1}\int_{\eta_{in}}^{\eta_{in}+T}d\eta\boldsymbol{p}_{x,y}(\eta)\exp[im\omega_{1}(\eta+2z)],
\end{equation}
from which one can see why the electron's transverse momenta $p_{x}$ and $p_{y}$ and longitudinal displacement $z$ are important here. The constant of motion is obtained as $\varsigma =\gamma-p_{z}=\gamma_{0}-p_{z0}$, where $\gamma$ is the electron relativistic factor derived from electron velocity, $p_{z}$ is the $z$ component of the electron relativistic momentum and $\gamma_{0},~p_{z0}$ are the initial value of them.
The fundamental frequency $\omega_1$ in dimensionless form for backscattering reads now
\begin{equation}\label{eqbasic3}
\omega_{1}=\frac{2\pi}{T+2z_{0}},
\end{equation}
where $T$ is the period of electronic periodic motion and ${\textbf{\textit{r}}}_{\textbf{0}}=(0,0,z_{0})$ is the drift distance.

The laser field is composed of two plane waves as $A_{1}$ and $A_{2}$. The ratio of the intensity is $\lambda$ and the ratio of the frequency is $\mu$. Since the laser propagation and external magnetic field are along in $z$ direction, the combinational total vector potential of combined fields can be written as
\begin{align}\label{eqe1}
  \textbf{\textit{A}}=&\textbf{\textit{A}}_{1}+\textbf{\textit{A}}_{2}+\textbf{\textit{A}}_{3}\nonumber\\
  =&\frac{A_{0}}{\sqrt{1+\alpha^{2}}} (-\sin\eta\hat{\textbf{i}}+\alpha\cos\eta\hat{\textbf{j}})\nonumber\\
  &+\frac{\lambda A_{0}}{\sqrt{1+\alpha^{2}}} (-\sin\mu\eta\hat{\textbf{i}}+\alpha\cos\mu\eta\hat{\textbf{j}})\nonumber\\
  &+B_{0}x\hat{\textbf{j}}£¬
\end{align}
where $\alpha$ stand for the ellipticity of the elliptically polarized plane wave. By the way it is noted that because there is no scalar potential two gauges, Coulomb one or Lorentz one, are equivalent to each other.

We normalize time by $1/\omega_{0}$, velocity by $c$, distance by $k_{0}^{-1}=c/\omega_{0}$, and momentum by $mc$. We obtain the equations for the momentum of the electron as following forms.
\begin{eqnarray}
\label{eqe2}
\frac{d^2 p_{x} }{d \eta ^2}+{\omega _{b}}^2 p_{x}=&
(\omega_{b} +\frac{1}{\sqrt{1+\alpha^{2}}}){a}\sin{\eta}+(\omega_{b} +\frac{\mu}{\sqrt{1+\alpha^{2}}}){a}\mu\lambda\sin{(\mu\eta)},\\
\label{eqc3}\frac{d^2 p_{y} }{d \eta ^2}+{\omega _{b}}^2 p_{y}=&
- (\omega_{b} +\frac{1}{\sqrt{1+\alpha^{2}}}){a}\cos{\eta }- (\omega_{b} +\frac{\mu}{\sqrt{1+\alpha^{2}}}){a}\mu\lambda\cos{(\mu\eta )},
\end{eqnarray}
where $\omega_{b}= b/ \varsigma$ with $a=eA_{1}/mc^2$ and $b=eB_{0}/m\omega_{0}c$ being the normalized vector potential and magnetic field, respectively.

The transverse momentum of the electron are obtained and the longitudinal momentum can also be obtained with the help of $p_{z}=\frac{p_{x}^{2}+p_{y}^{2}+1-\varsigma^{2}}{2\varsigma}$. Furthermore, assuming that at $t=0$ the electron is static and located at $x=0$, $y=0$ and $z=z_{in}$, so that $\eta=\eta_{in}=-z_{in}$. Moreover the momentum and longitudinal displacement of electron, $z_0$, would be easily got from the trajectory equations of the electron $d \boldsymbol{r}/d \eta=\boldsymbol{p}/\varsigma$. These expressions are not presented here since the formula are lengthy for a general elliptical case. We just point out a fact that $p_{x}$ and $p_{y}$ are linearly superimposed while $z$ is nonlinear, which will then lead to a great influence on the following study about the backscattering.

From now on for the simple and convenience calculations and discussions, we just consider circular case, i.e. ellipticity $\alpha=1$. In this case, the required momentum and displacement are expressed as
\begin{eqnarray}
\label{eqc2}
p_{x}=na\left \{ \sin{\eta}-\sin{\left[\omega_{b}\eta-\left( \omega_{b}-1\right)\eta_{in}\right]} \right \}
+la\left \{ \sin{\mu\eta}-\sin{\left[\omega_{b}\eta-\left( \omega_{b}-\mu\right)\eta_{in}\right]} \right \},\\
\label{eqc3}
p_{y}=na\left \{ \cos{\left[\omega_{b}\eta-\left( \omega_{b}-1\right)\eta_{in}\right]}-\cos{\eta} \right \}
+la\left \{ \cos{\left[\omega_{b}\eta-\left( \omega_{b}-\mu\right)\eta_{in}\right]}-\cos{\mu\eta} \right \},
\end{eqnarray}
\begin{align}\label{eqc4}
z(\eta)&=\left ( \frac{na}{\varsigma} \right )^{2}\left \{ \left ( \eta-\eta_{in} \right )-\frac{1}{\omega_{b}-1}\sin{\left [ \left ( \omega_{b}-1 \right )\left ( \eta-\eta_{in} \right ) \right ]} \right \}\nonumber\\
&+\left ( \frac{la}{\varsigma} \right )^{2}\left \{ \left ( \eta-\eta_{in} \right )-\frac{1}{\omega_{b}-\mu}\sin{\left [ \left ( \omega_{b}-\mu \right )\left ( \eta-\eta_{in} \right ) \right ]} \right \}\nonumber\\
&+\left ( \frac{1-\varsigma^2}{2\varsigma^2} \right )\left ( \eta-\eta_{in} \right ).
\end{align}
The trajectory equations of the electron can be obtained with the help of $p_{z}=\frac{p_{x}^{2}+p_{y}^{2}+1-\varsigma^{2}}{2\varsigma}$ and $d \boldsymbol{r}/d \eta=\boldsymbol{p}/\varsigma$.
Then the drift distance can be expressed as
\begin{equation}\label{eqc5}
z_{0}=\frac{2n\pi}{\varsigma^{2}}
\left(n^{2}a^{2}+l^{2}a^{2}+\frac{1}{2}-\frac{\varsigma^{2}}{2}\right),
\end{equation}
where $n=\frac{1}{\omega_{b}-1}$ and $l=\frac{\mu\lambda}{\omega_{b}-\mu}=\frac{n\lambda}{(n+1)/\mu-n}$.

Substituting Eqs.(\ref{eqc2}$-$\ref{eqc5}) into Eqs.~(\ref{eqbasic1}-\ref{eqbasic3}), we can calculate the backscattering spectra, on which the following numerical results are based.

By the similar discussion about the period $T=2\pi n$ as in Refs.\cite{Fu2016Scale,Jiang2017Thomson}, and keep in mind that $r_{0}= (0,0,z_{0})$ in Eq.(\ref{eqbasic3}), then we have the fundamental frequency
\begin{equation}\label{eq-frequency}
\omega_1 = \frac{\varsigma^{2}/n}{1+2a^{2}n^{2}+2a^{2}l^{2}}
=\frac{\varsigma^{2}/n}{1+2a^{2}n^{2}+2a^{2}[\frac{n\lambda}{(n+1)/\mu-n}]^2}.
\end{equation}
Obviously this fundamental frequency depends on parameters $a$, $n$, $p_{z0}$, $\lambda$, and $\mu$.

In order to qualitatively analysis and semi-quantitatively analyze the typical features of the Thomson backscatter spectrum, we can make an appropriate estimation to the emission intensity. For simplicity we only consider the case of $p_{z0}=0$, i.e., $\varsigma=1$.
Now the analytical expression of radiation spectrum is approximated as
\begin{equation}
\label{l1}\frac{d^2 I_{m}}{dt d\Omega} \approx \frac{e^2}{2\pi^2c}\left ( m\omega_{1} \right )^2\left (\pi \omega_{1}nla \right )^{2}
J^2_{0}(z_{m}),
\end{equation}
where $z_{m}=m\omega_{1} 2n^3a^2=\frac{2mn^3a^2}{n+2n^3a^2+2nl^2a^2}$ is defined to simplify the formula of radiation spectrum.
The asymptotic form of Bessel functions, $J_{\nu}(x) \sim\sqrt{\frac{2}{x\pi}} \cos(x-\frac{\pi}{2}\nu-\frac{\pi}{4})$ when $x\gg|\nu^{2}-1/4|$, obviously it is met since $z_m \sim m \gg1/4$, makes us to simply Eq.~(\ref{l1}) further as
\begin{equation}
\label{l2}\frac{d^2 I_{m}}{dt d\Omega}\approx\frac{e^2}{2\pi^2c}\left ( m\omega_{1} \right )^2\left (\pi \omega_{1}nla \right )^{2}J^2_{0}(z_{m})
\approx\frac{e^2}{4\pi^2c}y_{1}y_{2},
\end{equation}
with
\begin{equation}
\label{l3}
y_{1}=\frac{m\pi\omega_{1}^{3}l^{2}}{2n},
\end{equation}
and
\begin{equation}
\label{l4}
y_{2}=1+\sin(2z_{m}).
\end{equation}
Obviously $y_{1}$ is a power function depending on parameters $a$, $n$, $m$, $\lambda$, and $\mu$, and $y_{2}$ is an oscillation one.

\section{Numerical results and analysis}\label{numeric}

For the ignorance of the radiation backreaction \cite{Fu2016Scale}, as mentioned in Introduction, the chosen parameters need to satisfy $n^{4}a^{5}<10^{7}$ in the case of $p_{z0}=0$ and $p_{z0}^{2}n^{4}a^{5}<10^{6}$ in the case of $p_{z0}\neq0$. So in the following all chosen parameters are compelled to satisfy these requirements.

From Eq.(\ref{eqc4}) we see that the displacement of the election in $z$ direction is nonlinear. It greatly influences the backscattering radiation intensity in Eq.(\ref{eqbasic2}). While we have examined and calculated the velocity, the acceleration and also the trajectory of electron, these results are not presented with figure in this letter since the limited scope. However it should be pointed out that
in strong magnetic field, the axial motion in two lasers are much larger than that only linear addition of them by $A_{1}$ and $A_{2}$. These results confirm that the axial focusing effect on spiral movement of electrons is enhanced by larger magnetic field as well the strong nonlinear effect via relativistic charged particle dynamics.

\begin{figure}[htbp]\suppressfloats
\includegraphics[width=8cm]{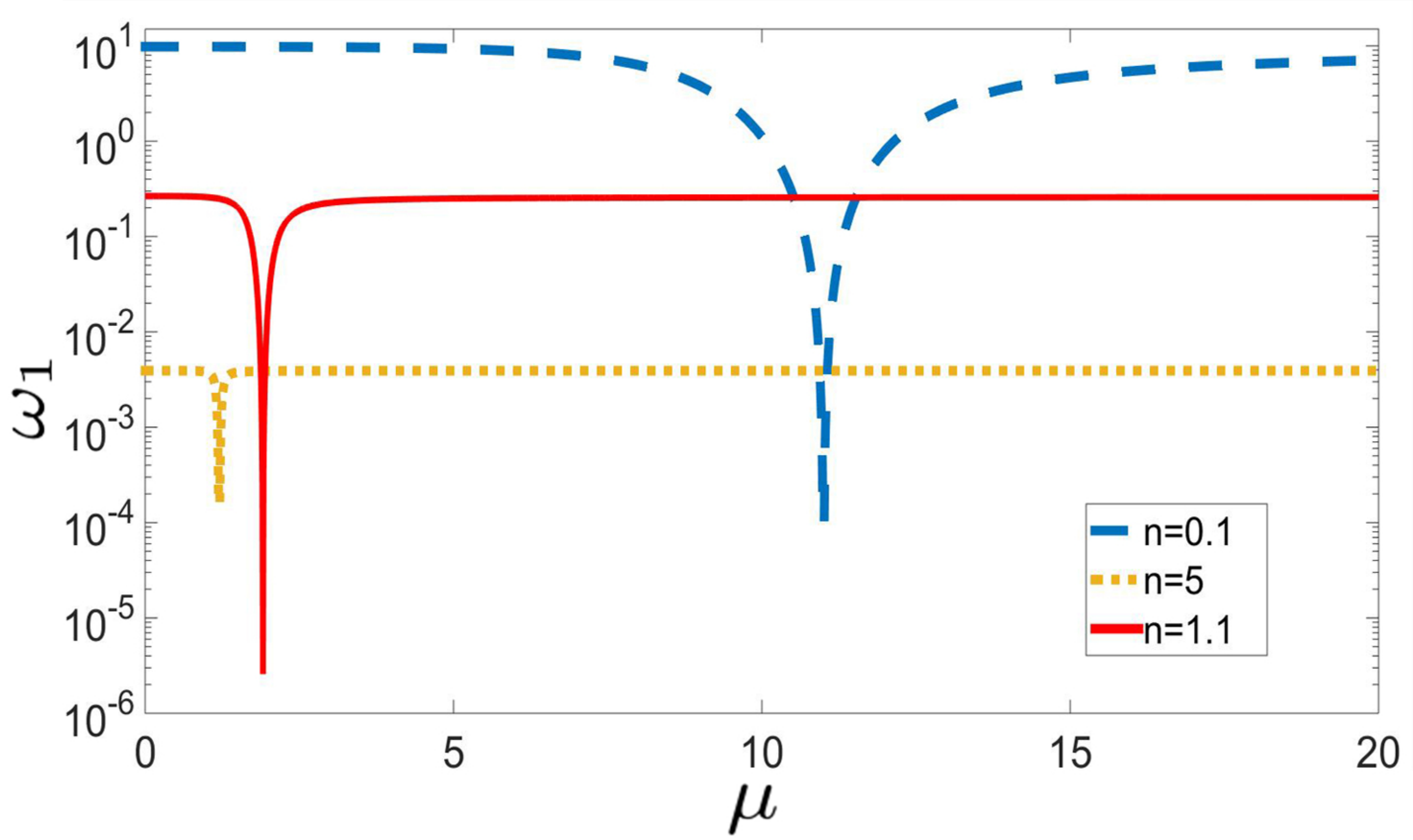}
\centering
\caption{\label{w1} Fundamental frequency $\omega_{1}$ depends on $\mu$ for different magnetic field cases of weak $n=1.1$ (red solid-line), resonance-like $n=5$ (yellow dotted-line) and strong $n=0.1$ (blue dashed-line), respectively. Other parameters are $a=1$, $\lambda=0.2$ and $p_{z0}=0$.}
\centering
\end{figure}

\begin{figure}[htbp]\suppressfloats
\includegraphics[width=13cm]{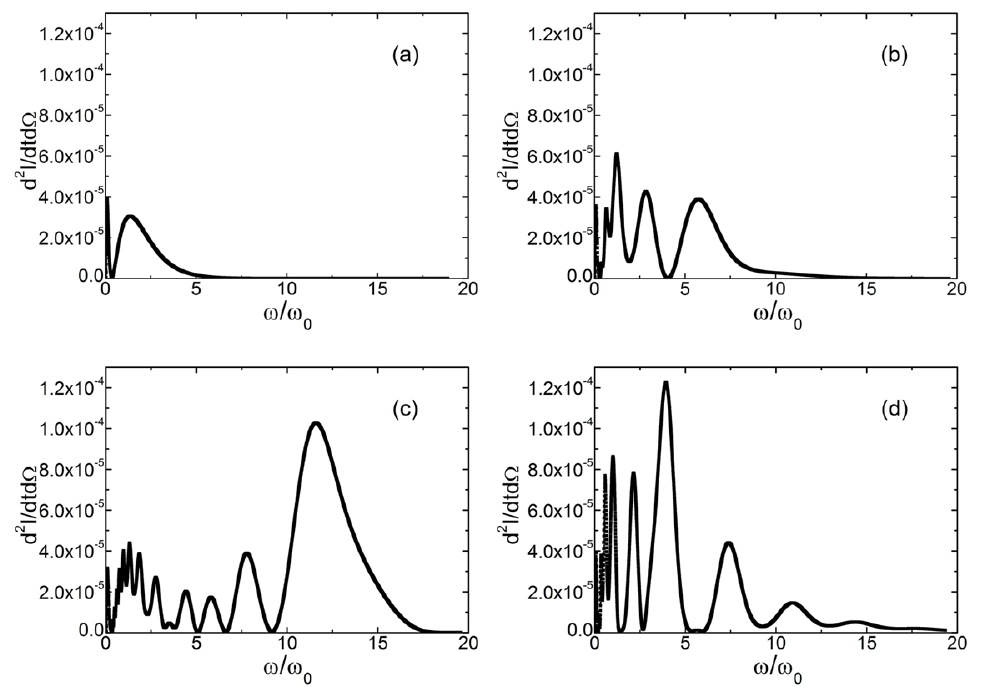}
\centering
\caption{\label{spectrum} The backscattering spectrum (normalized by $e^{2}/4\pi^{2}c$) for different laser cases of (a)$\lambda=0$; (b)$\lambda=0.2$, $\mu=8$; (c)$\lambda=0.2$, $\mu=15$ and (d)$\lambda=0.5$, $\mu=8$. The other parameters are $a=1$, $n=5$ and $p_{z0}=0$.}
\end{figure}

\begin{figure}[htbp]\suppressfloats
\includegraphics[width=14cm]{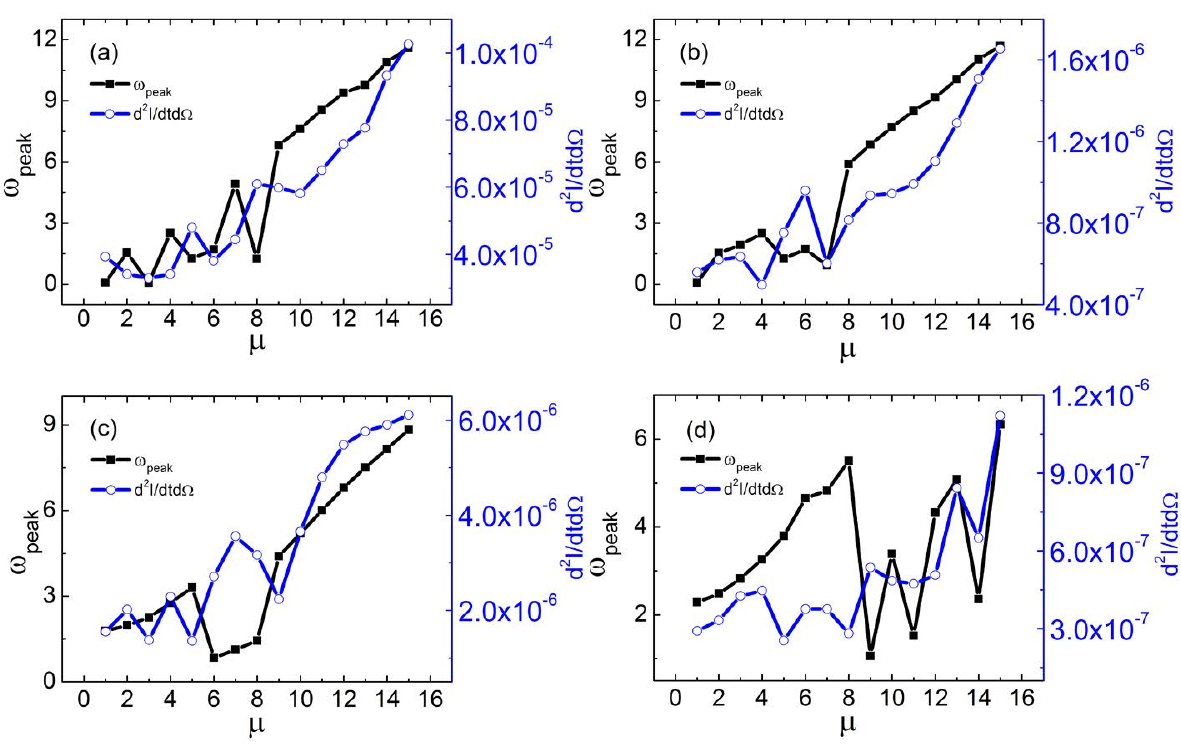}
\centering
\caption{\label{miu} The main-peak frequency of spectra (black-line with square-symbol) and the corresponding main-peak intensity normalized by $e^{2}/4\pi^{2}c$ (blue-line with circle-symbol) \emph{vs} two lasers frequency ratio, $\mu$, for different cases of (a)$a=1$, $n=5$; (b)$a=1$, $n=10$; (c)$a=2$, $n=5$ and (d)$a=3$, $n=5$. The other parameters are $\lambda=0.2$ and $p_{z0}=0$.}
\end{figure}

\begin{figure}[htbp]\suppressfloats
\includegraphics[width=14cm]{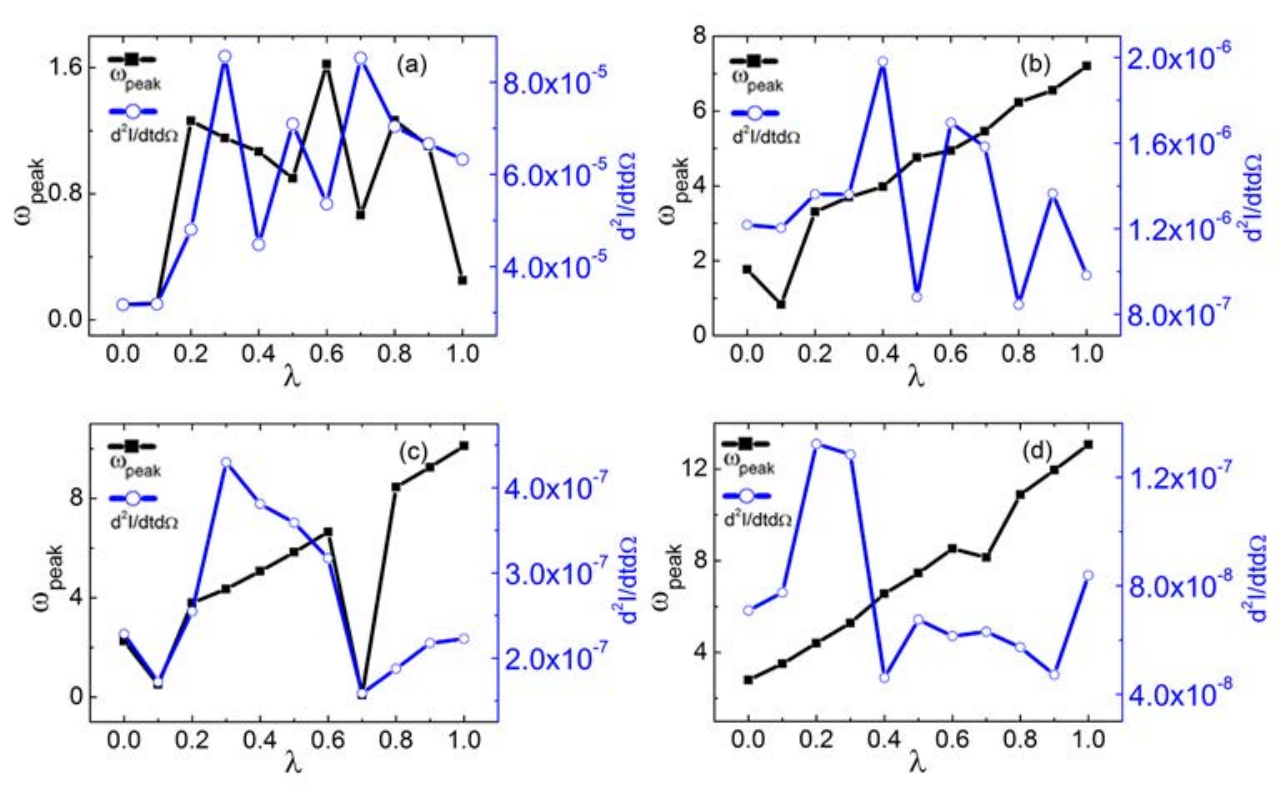}
\centering
\caption{\label{lamda} The main-peak frequency of spectra (black-line with square-symbol) and the corresponding main-peak intensity normalized by $e^{2}/4\pi^{2}c$ (blue-line with circle-symbol) \emph{vs} two lasers intensity ratio, $\lambda$, for different cases of  first laser field intensity: (a)$a=1$; (b)$a=2$; (c)$a=3$ and (d)$a=4$. The other parameters are $n=5$, $\mu=5$ and $p_{z0}=0$.}
\end{figure}

Since the fundamental frequency $\omega_{1}$ depends on parameters $a$, $n$, $p_{z0}$, $\lambda$, and $\mu$, we can briefly discuss the relation between these parameters and the fundamental frequency. First from expression Eq.(\ref{eq-frequency}), it is expected the results that $\omega_1$ decreases as $a$, $n$, $p_{z0}$, $\lambda$. In fact our numerical calculations have confirmed these conclusions. However since the $\mu$ can lead to the resonance case where $\mu_{res}=\omega_b=1+1/n$ which make the $l$ very large so that $\omega_1$ would be reduced drastically. This interesting nontrivial results are shown in Fig.~\ref{w1}. Indeed in the resonance case of second laser frequency matching to the cyclotron frequency, the fundamental frequency is decreased drastically, which is very favorite to THz regime.

Now let us turn to see the influence of two lasers on Thomson backscattering spectrum.
In Fig.~\ref{spectrum}, the backscattering spectrum (normalized by $e^{2}/4\pi^{2}c$) in four cases are shown. For simplicity the initial axial momentum $p_{z0}=0$ is considered and $a=1$, $n=5$. Other parameters are: (a)$\lambda=0$; (b)$\lambda=0.2$ and $\mu=8$; (c)$\lambda=0.2$ and $\mu=15$; (d)$\lambda=0.5$ and $\mu=8$. Obviously, Fig.~\ref{spectrum}(a) is the spectrum corresponds to case of only $A_{1}$. Compared to Fig.~\ref{spectrum}(a), the remaining three plotting show that the frequency and intensity of the spectrum tend to increase in the case of $A_{1}+A_{2}$, see Figs.~\ref{spectrum}(b), (c) and (d).

By comparing Fig.~\ref{spectrum} (c) with (b), one can see that the emission spectra moves to higher frequency regime meanwhile the high-frequency emission intensity is enhanced and the low-frequency emission is reduced as $A_{2}$ field frequency increases.
When the ratio of frequency of two lasers $\mu$ turns from $8$ into $15$, less than two times, however, the values of the harmonic orders $m$ changes from 318 to 2964, and the frequency where the main-peak emission locates is increased about ten times from $\omega / \omega_0\sim 1$ to $\sim12$. On the other hand the main-peak intensity is also increased about two times. It provides an alternative way to modulate and control the frequency and intensity of backscattering spectra by adjusting the second weak laser field frequency in addition to the first strong laser field.

Comparing Figs.~\ref{spectrum} (d) with (b), when the intensity of $A_{2}$ increases and other parameters are fixed, it is not obvious that the spectra move to high frequency regime as well as the main-peak frequency shift, while the main-peak intensity is increased about two times at the same order of the $A_2$ intensity enlargement.

It is worthy to note that, in our studied two-laser case, the main peak of the radiation spectra no longer appears at the fundamental frequency like in one-laser case of Fig.~\ref{spectrum} (a). With different parameters, main peak will appear in low-frequency [(b)], high-frequency [(c)] or intermediate [(d)] region of the radiation spectrum. It also implies that the addition of $A_{2}$ to $A_{1}$ will change the shape of the spectrum nonlinearly.

The dependence of the main-peak frequency (black with square) in spectra and corresponding emission intensity (blue with circle) on the ratio of two lasers' frequency, $\mu$, is shown in Fig.~\ref{miu}. The parameters are chosen as $p_{z0}=0$, $\lambda=0.2$ and (a): $a=1$, $n=5$; (b): $a=1$, $n=10$; (c)$a=2$, $n=5$; (d)$a=3$, $n=5$. It is found that
the frequency at peak, $\omega_{peak}=m\omega_1$, and corresponding intensity both tend to increase with $\mu$. In addition, this increasing tendency is not a linear relation but with an oscillatory increasing behavior. Because the peak is probably in the low-frequency region or the high-frequency region of the radiation spectrum rather than only fundamental frequency, as mentioned above and observed in Fig~\ref{spectrum}. It reflects indeed the nonlinear effects of two lasers on Thomson backscattering.

Similarly we can also see the dependence of the main-peak frequency (black with square) of the spectra and corresponding emission intensity (blue with circle) on the ratio of two lasers' intensity, $\lambda$, which are shown in Fig.~\ref{lamda}. Now $p_{z0}=0$, $\mu=5$ and $n=5$ are fixed but the different laser intensity of $A_1$ are given as $a=1,~2,~3,~4$ for (a)-(d), respectively. It is found that
as the intensity of $A_{1}$ increases, obviously the dependence of the main-peak frequency on $\lambda$ seems to increase nevertheless also in an oscillation way. However, the corresponding peak intensity is changing with $\lambda$ either ascendant or descendant, which exhibits so high nonlinear characteristic that there is no obvious global tendency.

Here we maybe have a simple but qualitative discussion on the above results. In fact as is shown in Eqs.~(\ref{l2}-\ref{l4}), $y_{2}$ oscillates as $\mu$ or $\lambda$ increases, so we pay attention to the value of $y_{1}$. When $\mu$ increases, in particular in most cases of $\mu>>\mu_{res}=1+1/n$, the parameter $l$ and $\omega_{1}$ almost does not change with $\mu$, refer to Fig.~\ref{w1}. Therefore as $\omega_{peak}=m \omega_{1}$ increases as $m$, the radiation intensity increases too but in oscillatory way. On the other hand, when $\lambda$ increases, the three parameters $l$, $\omega_{1}$ and $m$ are all changed, which affect the $y_1$ in a complex nonlinear way. So that the strong nonlinear characteristics of the intensity are exhibited in Fig.~\ref{lamda}
Although they do not show a strict relation, we can adjust the backscattering spectra to what we need by adjusting the appropriate intensity ratio of two lasers. For example, in the Fig.~\ref{lamda}(b), by slightly increasing the intensity of the $A_{2}$ from $0.5$ to $0.6$, the backscattering spectrum increases by about two times in the peak intensity, however, the corresponding peak frequency is shifted little.

\section{Conclusions and discussions}\label{conclusion}

In this Letter, we study the Thomson backscattering of an electron in combined two lasers and constant magnetic fields with dynamically assisted mechanism. The emission fundamental frequency and backscattering spectra are obtained and then analyzed. We have revealed the effects of some factors on the Thomson backscattering.

In summary. The nonlinear effects of axial motion of the electron would result in nontrivial behavior on Thomson backscatter spectra. When the second laser frequency matching to the cyclotron frequency, the decreased fundamental frequency is favorite to THz regime. Moreover by changing the frequency or the intensity ratio of high-frequency to low-frequency lasers, the main-peak frequency of the backscattering spectrum and the intensity of the peak intensity exhibits strong nonlinear characteristics.

It should be pointed out that the dynamically assisted mechanism is studied extensively in the research of pair production in strong background fields, however, it is the first time to be applicable to the Thomson scattering study presented here. The enhanced spectra intensity, the high nonlinear characteristics as well as the controllability of emitted spectral bandwidth are evident by this mechanism.

On the other hand, the present research topic has still some open problems. For example,
we only discuss backscattering under two laser fields and a magnetic field, while scattering in other directions are interesting and important, as we did before with respect to forward and full-angle scattering under a single laser field\cite{Zhao2018Angular,Huang2018Effects}.
Another very interesting and invaluable point is to consider the more realistic laser with profile and electron bunch with initial distribution. Moreover, the concrete interference structure due to two fields' nonlinear addition of the corresponding electron's displacement, which appears first in the phase part of Thomson scattering formula, i.e. the exponent function part of Eq.(\ref{eqbasic2}) and then the modulus as in Eq.(\ref{eqbasic1}), and how this interference affects the final spectra are still open. However these complicated and abundant topics exceed the scope of present letter and certainly need to be researched in detail in future.

Anyway our results presented here provide a new possibility to adjust or/and control the Thomson spectra for the THz or/and x-ray light source. Obviously, the production of either THz or x-ray depends on the fundamental frequency which is strongly associated to the electrons' initial longitudinal momentum, two lasers' parameters as well as the applied magnetic field.

\begin{acknowledgments}
First author are thankful to J. Huang and Dr. C. Lv for useful discussions. This work was supported by the National Natural Science Foundation of China (NSFC) under Grant No.11875007 and No.11305010. The computation was carried out at the HSCC of the Beijing Normal University.
\end{acknowledgments}

\end{document}